\documentclass[12pt,a4paper]{article}

\usepackage{ifthen} 
\newboolean{pdflatex}
\setboolean{pdflatex}{true} 

\newboolean{articletitles}
\setboolean{articletitles}{true} 

\newboolean{uprightparticles}
\setboolean{uprightparticles}{false} 

\newboolean{wordcount}
\setboolean{wordcount}{false} 

\ifthenelse{\boolean{wordcount}}{ 

\usepackage{xspace} 
\usepackage{upgreek}

\def\lhcb {\mbox{LHCb}\xspace}

\def\dzero  {\mbox{D0}\xspace}

\def\MagUp {\mbox{\em Mag\kern -0.05em Up}\xspace}

\ifthenelse{\boolean{uprightparticles}}%
{

 \def\Pmu         {\ensuremath{\upmu}\xspace}

 \def\Ppi         {\ensuremath{\uppi}\xspace}

 \def\Ppsi        {\ensuremath{\uppsi}\xspace}

 \def\PDelta      {\ensuremath{\Delta}\xspace}                 
 \def\PXi      {\ensuremath{\Xi}\xspace}                 
 \def\PLambda      {\ensuremath{\Lambda}\xspace}                 
 \def\PSigma      {\ensuremath{\Sigma}\xspace}                 
 \def\POmega      {\ensuremath{\Omega}\xspace}                 
 \def\PUpsilon      {\ensuremath{\Upsilon}\xspace}                 
 
 \def\PB      {\ensuremath{\mathrm{B}}\xspace}                 
                  
 \def\PD      {\ensuremath{\mathrm{D}}\xspace}

 \def\PJ      {\ensuremath{\mathrm{J}}\xspace}                 
 \def\PK      {\ensuremath{\mathrm{K}}\xspace}

 \def\Pb      {\ensuremath{\mathrm{b}}\xspace}                 
 \def\Pc      {\ensuremath{\mathrm{c}}\xspace}

 \def\Pi      {\ensuremath{\mathrm{i}}\xspace}

 \def\Pp      {\ensuremath{\mathrm{p}}\xspace}

 \def\Ps      {\ensuremath{\mathrm{s}}\xspace}

}
{

 \def\Pmu         {\ensuremath{\mu}\xspace}

 \def\Ppi         {\ensuremath{\pi}\xspace}

 \def\Ppsi        {\ensuremath{\psi}\xspace}                 
                  
 \mathchardef\PDelta="7101
 \mathchardef\PXi="7104
 \mathchardef\PLambda="7103
 \mathchardef\PSigma="7106
 \mathchardef\POmega="710A
 \mathchardef\PUpsilon="7107
                  
 \def\PB      {\ensuremath{B}\xspace}                 
                  
 \def\PD      {\ensuremath{D}\xspace}

 \def\PJ      {\ensuremath{J}\xspace}                 
 \def\PK      {\ensuremath{K}\xspace}

 \def\Pb      {\ensuremath{b}\xspace}                 
 \def\Pc      {\ensuremath{c}\xspace}

 \def\Pi      {\ensuremath{i}\xspace}

 \def\Pp      {\ensuremath{p}\xspace}

 \def\Ps      {\ensuremath{s}\xspace}

}

\makeatletter
\ifcase \@ptsize \relax
  \newcommand{\miniscule}{\@setfontsize\miniscule{4}{5}}
\or
  \newcommand{\miniscule}{\@setfontsize\miniscule{5}{6}}
\or
  \newcommand{\miniscule}{\@setfontsize\miniscule{5}{6}}
\fi
\makeatother

\DeclareRobustCommand{\optbar}[1]{\shortstack{{\miniscule (\rule[.5ex]{1.25em}{.18mm})}
  \\ [-.7ex] $#1$}}

\def\mumu       {{\ensuremath{\Pmu^+\Pmu^-}}\xspace}

\def\squark    {{\ensuremath{\Ps}}\xspace}

\def\cquark    {{\ensuremath{\Pc}}\xspace}

\def\bquark    {{\ensuremath{\Pb}}\xspace}

\def\pion   {{\ensuremath{\Ppi}}\xspace}

\def\pip    {{\ensuremath{\pion^+}}\xspace}
\def\pim    {{\ensuremath{\pion^-}}\xspace}
\def\pipm   {{\ensuremath{\pion^\pm}}\xspace}

\def\kaon    {{\ensuremath{\PK}}\xspace}
  \def\Kbar    {{\kern 0.2em\overline{\kern -0.2em \PK}{}}\xspace}

\def\KorKbar    {\kern 0.18em\optbar{\kern -0.18em K}{}\xspace}

\def\Kp      {{\ensuremath{\kaon^+}}\xspace}
\def\Km      {{\ensuremath{\kaon^-}}\xspace}
\def\Kpm     {{\ensuremath{\kaon^\pm}}\xspace}

  \def\Dbar    {{\kern 0.2em\overline{\kern -0.2em \PD}{}}\xspace}
\def\D       {{\ensuremath{\PD}}\xspace}

\def\DorDbar    {\kern 0.18em\optbar{\kern -0.18em D}{}\xspace}

\def\Dm      {{\ensuremath{\D^-}}\xspace}

\def\Dsm     {{\ensuremath{\D^-_\squark}}\xspace}

\def\Dsmp    {{\ensuremath{\D^{\mp}_\squark}}\xspace}

\def\B       {{\ensuremath{\PB}}\xspace}
\def\Bbar    {{\ensuremath{\kern 0.18em\overline{\kern -0.18em \PB}{}}}\xspace}

\def\BorBbar    {\kern 0.18em\optbar{\kern -0.18em B}{}\xspace}
\def\Bz      {{\ensuremath{\B^0}}\xspace}

\def\Bu      {{\ensuremath{\B^+}}\xspace}

\def\Bp      {{\ensuremath{\Bu}}\xspace}

\def\Bs      {{\ensuremath{\B^0_\squark}}\xspace}

\def\Bc      {{\ensuremath{\B_\cquark^+}}\xspace}

\def\jpsi     {{\ensuremath{{\PJ\mskip -3mu/\mskip -2mu\Ppsi\mskip 2mu}}}\xspace}

  \def\Y#1S{\ensuremath{\PUpsilon{(#1S)}}\xspace}

\def\proton      {{\ensuremath{\Pp}}\xspace}

\def\Lz          {{\ensuremath{\PLambda}}\xspace}
\def\Lbar        {{\ensuremath{\kern 0.1em\overline{\kern -0.1em\PLambda}}}\xspace}
\def\LorLbar    {\kern 0.18em\optbar{\kern -0.18em \PLambda}{}\xspace}

\def\Lb      {{\ensuremath{\Lz^0_\bquark}}\xspace}

\def\to                 {\ensuremath{\rightarrow}\xspace}

\def\AT#1     {\ensuremath{A_{\mathrm{T}}^{#1}}\xspace}           

\def\C#1      {\ensuremath{\mathcal{C}_{#1}}\xspace}                       
\def\Cp#1     {\ensuremath{\mathcal{C}_{#1}^{'}}\xspace}                    
\def\Ceff#1   {\ensuremath{\mathcal{C}_{#1}^{\mathrm{(eff)}}}\xspace}        
\def\Cpeff#1  {\ensuremath{\mathcal{C}_{#1}^{'\mathrm{(eff)}}}\xspace}       
\def\Ope#1    {\ensuremath{\mathcal{O}_{#1}}\xspace}                       
\def\Opep#1   {\ensuremath{\mathcal{O}_{#1}^{'}}\xspace}                    

\newcommand{\tev}{\ensuremath{\mathrm{\,Te\kern -0.1em V}}\xspace}
\newcommand{\gev}{\ensuremath{\mathrm{\,Ge\kern -0.1em V}}\xspace}
\newcommand{\mev}{\ensuremath{\mathrm{\,Me\kern -0.1em V}}\xspace}
\newcommand{\kev}{\ensuremath{\mathrm{\,ke\kern -0.1em V}}\xspace}
\newcommand{\ev}{\ensuremath{\mathrm{\,e\kern -0.1em V}}\xspace}
\newcommand{\gevc}{\ensuremath{{\mathrm{\,Ge\kern -0.1em V\!/}c}}\xspace}
\newcommand{\mevc}{\ensuremath{{\mathrm{\,Me\kern -0.1em V\!/}c}}\xspace}
\newcommand{\gevcc}{\ensuremath{{\mathrm{\,Ge\kern -0.1em V\!/}c^2}}\xspace}
\newcommand{\gevgevcccc}{\ensuremath{{\mathrm{\,Ge\kern -0.1em V^2\!/}c^4}}\xspace}
\newcommand{\mevcc}{\ensuremath{{\mathrm{\,Me\kern -0.1em V\!/}c^2}}\xspace}
\newcommand{\nspmevcc}{\ensuremath{{\mathrm{Me\kern -0.1em V\!/}c^2}}\xspace}

\def\mum  {\ensuremath{{\,\upmu\mathrm{m}}}\xspace}

\def\invfb   {\ensuremath{\mbox{\,fb}^{-1}}\xspace}

\newcommand{\stat}{\ensuremath{\mathrm{\,(stat)}}\xspace}
\newcommand{\syst}{\ensuremath{\mathrm{\,(syst)}}\xspace}

\def\gsim{{~\raise.15em\hbox{$>$}\kern-.85em
          \lower.35em\hbox{$\sim$}~}\xspace}
\def\lsim{{~\raise.15em\hbox{$<$}\kern-.85em
          \lower.35em\hbox{$\sim$}~}\xspace}

\def\ptot       {\mbox{$p$}\xspace}
\def\pt         {\mbox{$p_{\mathrm{ T}}$}\xspace}

\def\evtgen     {\mbox{\textsc{EvtGen}}\xspace}

\def\geant      {\mbox{\textsc{Geant4}}\xspace}

\def\photos     {\mbox{\textsc{Photos}}\xspace}

\def\pythia     {\mbox{\textsc{Pythia}}\xspace}

\def\tell1  {TELL1\xspace}
\def\ukl1   {UKL1\xspace}

\def\Bss      {{\ensuremath{\B^{*0}_\squark}}\xspace}

  \usepackage{bibentry} 
  \usepackage{mciteplus} 
  \usepackage{comment} 
  \excludecomment{align*} 
  \excludecomment{align} 
  \excludecomment{equation*} 
  \excludecomment{equation} 
  \excludecomment{eqnarray*} 
  \excludecomment{eqnarray} 
}{
}
           
\ifthenelse{\boolean{wordcount}}{}{
  
\usepackage[top=1in, bottom=1.25in, left=1in, right=1in]{geometry}

\columnsep=5mm
\addtolength{\belowcaptionskip}{0.5em}

\raggedbottom
\sloppy

\usepackage{microtype}
\usepackage{lineno}  
\usepackage{xspace} 
\usepackage{caption} 

\usepackage{graphicx}  
\usepackage{color}
\usepackage{colortbl}
\graphicspath{{./figs/}} 

\usepackage{amsmath} 
\usepackage{amssymb}
\usepackage{amsfonts}
\usepackage{upgreek} 

\newcommand*\patchAmsMathEnvironmentForLineno[1]{%
\expandafter\let\csname old#1\expandafter\endcsname\csname #1\endcsname
\expandafter\let\csname oldend#1\expandafter\endcsname\csname
end#1\endcsname
 \renewenvironment{#1}%
   {\linenomath\csname old#1\endcsname}%
   {\csname oldend#1\endcsname\endlinenomath}%
}
\newcommand*\patchBothAmsMathEnvironmentsForLineno[1]{%
  \patchAmsMathEnvironmentForLineno{#1}%
  \patchAmsMathEnvironmentForLineno{#1*}%
}
\AtBeginDocument{%
\patchBothAmsMathEnvironmentsForLineno{equation}%
\patchBothAmsMathEnvironmentsForLineno{align}%
\patchBothAmsMathEnvironmentsForLineno{flalign}%
\patchBothAmsMathEnvironmentsForLineno{alignat}%
\patchBothAmsMathEnvironmentsForLineno{gather}%
\patchBothAmsMathEnvironmentsForLineno{multline}%
\patchBothAmsMathEnvironmentsForLineno{eqnarray}%
}

\usepackage{hyperref}    
\usepackage[all]{hypcap} 

\usepackage{cite} 
\usepackage{mciteplus}

}

\begin{document}

\renewcommand{\thefootnote}{\fnsymbol{footnote}}
\setcounter{footnote}{1}



\begin{titlepage}
\pagenumbering{roman}

\vspace*{-1.5cm}
\centerline{\large EUROPEAN ORGANIZATION FOR NUCLEAR RESEARCH (CERN)}
\vspace*{1.5cm}
\noindent
\begin{tabular*}{\linewidth}{lc@{\extracolsep{\fill}}r@{\extracolsep{0pt}}}
\ifthenelse{\boolean{pdflatex}}
 & &\\
 & & CERN-EP-2016-189 \\  
 & & LHCb-PAPER-2016-029 \\  
 & & 6 September 2016 \\ 
 & & \\
\end{tabular*}

\vspace*{4.0cm}

{\normalfont\bfseries\boldmath\huge
\begin{center}
  Search for structure in the $\Bs\pipm$ invariant mass spectrum
\end{center}
}

\vspace*{2.0cm}

\begin{center}
The LHCb collaboration\footnote{Authors are listed at the end of this Letter.}
\end{center}

\vspace{\fill}

\begin{abstract}
  \noindent
  The $\Bs\pipm$ invariant mass distribution is investigated in order to search for possible exotic meson states.
  The analysis is based on a data sample recorded with the LHCb detector corresponding to $3 \invfb$ of $pp$ collision data at $\sqrt{s} = 7$ and $8 \tev$.
  No significant excess is found, and upper limits are set on the production rate of the claimed $X(5568)$ state within the LHCb acceptance.
  Upper limits are also set as a function of the mass and width
  of a possible exotic meson decaying to the $\Bs\pipm$ final state. The same limits also apply to a possible exotic meson decaying through the chain $\Bss\pipm$, $\Bss \to \Bs \gamma$ where the photon is excluded from the reconstructed decays.
\end{abstract}

\vspace*{2.0cm}

\begin{center}
  Published in Phys.~Rev.~Lett. 117 (2016) 152003
\end{center}

\vspace{\fill}

{\footnotesize 
\centerline{\copyright~CERN on behalf of the \lhcb collaboration, licence \href{http://creativecommons.org/licenses/by/4.0/}{CC-BY-4.0}.}}
\vspace*{2mm}

\end{titlepage}


\newpage
\setcounter{page}{2}
\mbox{~}

\cleardoublepage


\renewcommand{\thefootnote}{\arabic{footnote}}
\setcounter{footnote}{0}


\pagestyle{plain} 
\setcounter{page}{1}
\pagenumbering{arabic}

Interest in exotic hadrons has recently intensified, with a wealth of experimental data becoming available~\cite{Olsen:2014qna,Cheng:2015vpn}.
All the well-established exotic states contain a heavy quark-antiquark ($c\bar{c}$ or $b\bar{b}$) pair together with additional light particle content.
However, the \dzero collaboration has reported evidence~\cite{D0:2016mwd} of a narrow structure, referred to as the $X(5568)$, in the $\Bs\pipm$ spectrum produced in $p\bar{p}$ collisions at centre-of-mass energy $\sqrt{s} = 1.96 \tev$.
The claimed $X(5568)$ state, if confirmed, would differ from any of the previous observations, as it must have constituent quarks with four different flavours \mbox{($b$, $s$, $u$, $d$)}.
As such, it would be unique among observed exotic hadrons in having its mass dominated by a single constituent quark rather than by a quark-antiquark pair.
This could provide a crucial piece of information to help understand how exotic hadrons are bound; specifically, whether they are dominantly tightly bound (often referred to as ``tetraquarks'' and ``pentaquarks'') or loosely bound meson-meson or meson-baryon molecules.

In this Letter, results are presented from a search for an exotic meson, denoted $X$, decaying to $\Bs\pipm$ in a data sample corresponding to $3\invfb$ of $pp$ collision data at $\sqrt{s} = 7$ and $8 \tev$ recorded by LHCb.
The search is performed by scanning over the mass and width of the purported state, with dedicated fits for parameters corresponding to those of the claimed $X(5568)$ state.
The $\Bs$ mesons are reconstructed in decays to $\Dsm\pip$ and $\jpsi\phi$ final states to obtain a \Bs yield approximately 20 times larger than that used by the \dzero collaboration.
The inclusion of charge-conjugate processes is implied throughout the Letter.
The analysis techniques follow closely those developed for studies of the $\Bp\Km$~\cite{LHCb-PAPER-2012-030}, $\Bp\pim$ and $\Bz\pip$~\cite{LHCb-PAPER-2014-067} spectra.
As in previous analyses, the charged pion which is combined with the \Bs meson in order to form the $\Bs\pipm$ candidate is referred to as the ``companion pion''.

The \lhcb detector~\cite{Alves:2008zz,LHCb-DP-2014-002} is a single-arm forward spectrometer covering the \mbox{pseudorapidity} range $2<\eta <5$, designed for the study of particles containing \bquark or \cquark quarks. 
The detector includes a high-precision tracking system consisting of a silicon-strip vertex detector surrounding the $pp$ interaction region, a large-area silicon-strip detector located upstream of a dipole magnet with a bending power of about $4{\mathrm{\,Tm}}$, and three stations of silicon-strip detectors and straw drift tubes placed downstream of the magnet.
The tracking system provides a measurement of momentum, \ptot, of charged particles with a relative uncertainty that varies from $0.5\,\%$ at low momentum to $1.0\,\%$ at $200\gev$ (units in which $c = \hbar = 1$ are used throughout).
The minimum distance of a track to a primary vertex (PV), the impact parameter, is measured with a resolution of $(15+29/\pt)\mum$, where \pt is the component of the momentum transverse to the beam, in~\gev.
Different types of charged hadrons are distinguished using information from two ring-imaging Cherenkov detectors. 
Photons, electrons and hadrons are identified by a calorimeter system consisting of scintillating-pad and preshower detectors, an electromagnetic calorimeter and a hadronic calorimeter. 
Muons are identified by a system composed of alternating layers of iron and multiwire proportional chambers.
The online event selection is performed by a trigger, which consists of a hardware stage, based on information from the calorimeter and muon systems, followed by a software stage, which applies a full event reconstruction.

Simulations of $pp$ collisions are generated using \pythia~\cite{Sjostrand:2006za,*Sjostrand:2007gs} with a specific \lhcb configuration~\cite{LHCb-PROC-2010-056}.  
Decays of hadronic particles are described by \evtgen~\cite{Lange:2001uf}, in which final-state radiation is generated using \photos~\cite{Golonka:2005pn}. 
The interaction of the generated particles with the detector, and its response, are implemented using the \geant toolkit~\cite{Allison:2006ve,*Agostinelli:2002hh} as described in Ref.~\cite{LHCb-PROC-2011-006}.

Candidate $\Bs$ mesons are reconstructed through the decays $\Bs\to\Dsm\pip$ with \mbox{$\Dsm\to\Kp\Km\pim$}, and $\Bs\to\jpsi\phi$ with $\jpsi \to\mumu$ and $\phi\to\Kp\Km$.
Particle identification, track quality and impact parameter requirements are imposed on all final-state particles.
Both \Bs and intermediate particle ($\Dsm$ and $\jpsi$) candidates are required to have good vertex quality and to have invariant mass close to the known values~\cite{PDG2014}.  
Specific backgrounds due to other \bquark-hadron decays are removed with appropriate vetoes.
A requirement is imposed on the multiplicity of tracks originating from the PV associated with the \Bs candidate; this requirement is about $90\,\%$ efficient on \Bs signal and significantly reduces background due to random $\Bs\pipm$ combinations.
To further reduce background, the \pt of the \Bs candidate, $\pt(\Bs)$, is required to be greater than $5 \gev$.
Results are also obtained with this requirement increased to $10$ or $15 \gev$, to be more sensitive to scenarios in which the $X$ state is predominantly produced from hard processes. 
The definition of the fiducial acceptance is completed with the requirements $\pt(\Bs) < 50 \gev$ and $2.0 < y < 4.5$, where $y$ is the rapidity of the \Bs candidate.

\begin{figure} [!tb]
  \centering
  \includegraphics[width=0.4\textwidth]{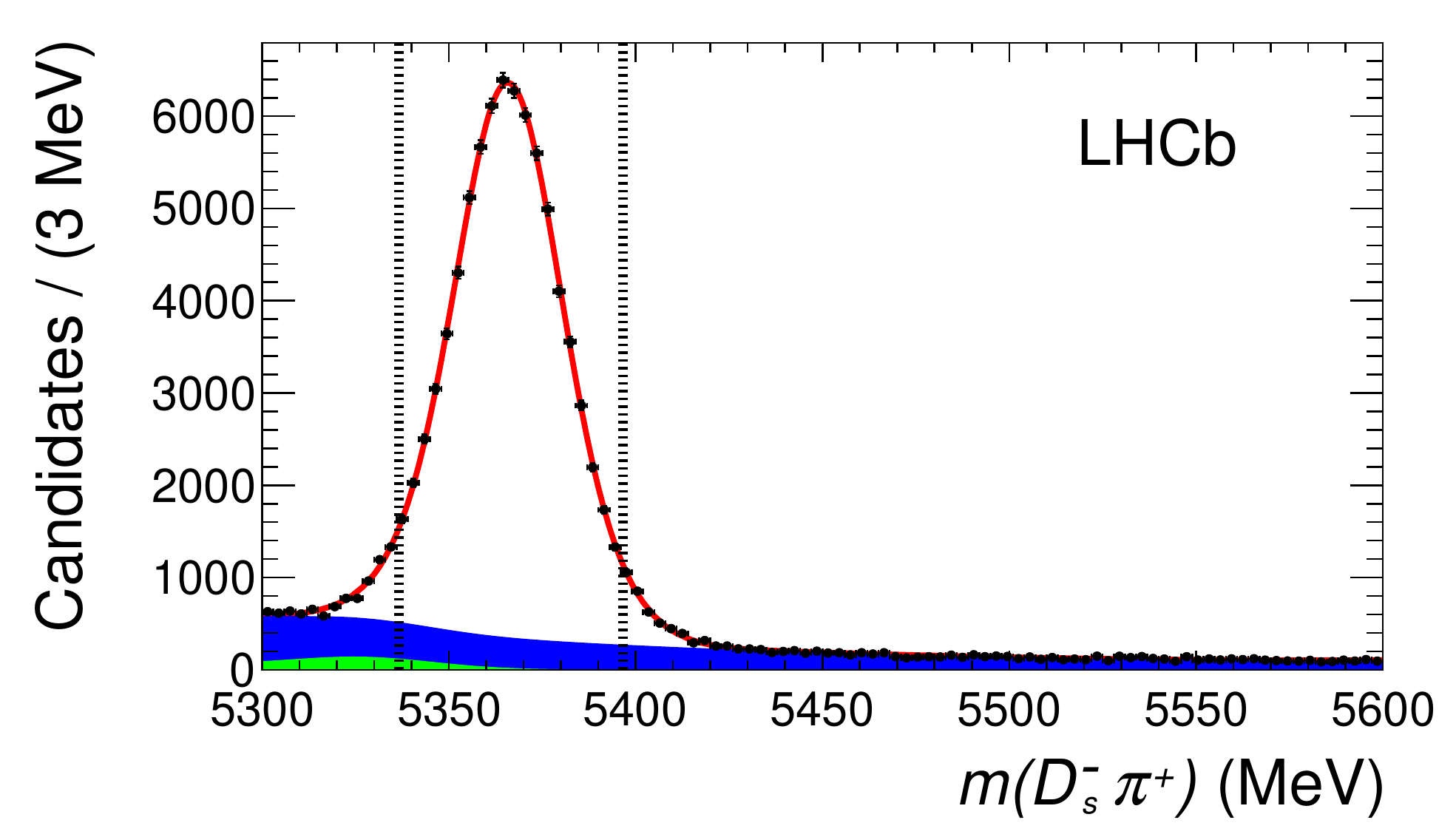}
  \includegraphics[width=0.4\textwidth]{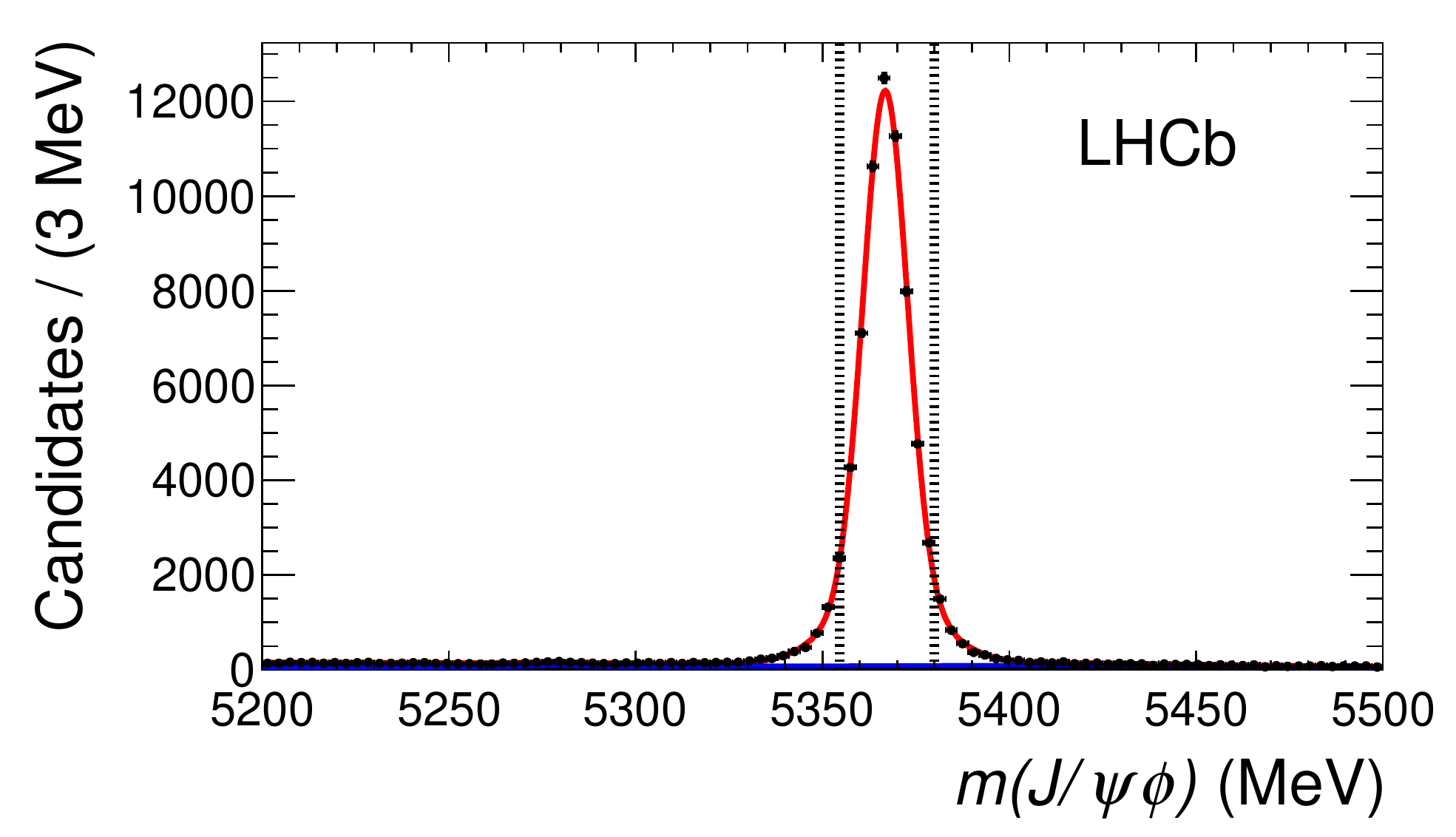}
  \caption{\small
    Selected candidates for (left) $\Bs \to \Dsm\pip$ and (right) $\Bs \to \jpsi\phi$ decays, with $\pt(\Bs) > 5 \gev$, where the \Bs signal window requirements of $|m(\Dsm\pip) - 5367 \mev| < 30 \mev$ and $|m(\jpsi\phi) - 5367 \mev| < 13 \mev$ are indicated by dotted lines.
    Results of the fits described in the text are superimposed with the total fit result shown as a red line, the signal component as an unfilled area, the combinatorial background component as a dark blue area and additional background contributions as a light green area.
  } 
  \label{fig:Bsignals}
\end{figure}

The signals in the two \Bs decay modes are shown in Fig.~\ref{fig:Bsignals}.
To estimate the \Bs yields, the data are fitted with functions that include a signal component, described by a double Gaussian function with a shared mean, and a combinatorial background component, described by a polynomial function.
Backgrounds from $\Bs \to \Dsmp\Kpm$ decays in the $\Dsm\pip$ sample and from
$\Lb \to \jpsi \proton\Km$ decays in the $\jpsi\phi$ sample, where a final-state hadron is misidentified, are modelled using empirical shapes derived from simulated samples.
An additional component, modelled with a Gaussian function, is included to account for possible $\Bz\to \jpsi\Kp\Km$ decays~\cite{LHCb-PAPER-2012-040} in the $\jpsi\phi$ sample.
The results of these fits are reported in Table~\ref{tab:fitResults}.
The signal-to-background ratio in the \Bs signal windows is about 10 for the $\Dsm\pip$ sample and above 50 for the $\jpsi\phi$ sample.

The \Bs candidates are combined with each track originating  from the associated PV that gives a good quality $\Bs\pipm$ vertex and that has $\pt > 500 \mev$.
A loose pion identification requirement is imposed in order to suppress possible backgrounds involving misidentified particles.
In case multiple candidates are obtained in the same event, all are retained.
Mass and vertex constraints are imposed~\cite{Hulsbergen:2005pu} in the calculation of the $\Bs\pipm$ invariant mass.

In order to obtain quantitative results on the contributions from resonant structures in the data, the $\Bs\pipm$ mass distributions are fitted with a function containing components for signal and background.
The signal shape is an S-wave Breit--Wigner function multiplied by a function that accounts for the variation of the efficiency with $\Bs\pipm$ mass.
The efficiency function, determined from simulation, plateaus at high $\Bs\pipm$ mass and falls near the threshold to a value that depends on $\pt(\Bs)$.
The resolution is better than $1\mev$ and does not affect the results.
The background is modelled with a polynomial function.
It is verified that this function gives a good description of backgrounds composed of either a real or a fake \Bs decay combined with a random pion, as determined from simulation or from data in \Bs candidate mass sideband regions, respectively.

\begin{figure}[!tb]
\ifthenelse{\boolean{wordcount}}{}{
  \centering
  \includegraphics[width=0.65\textwidth]{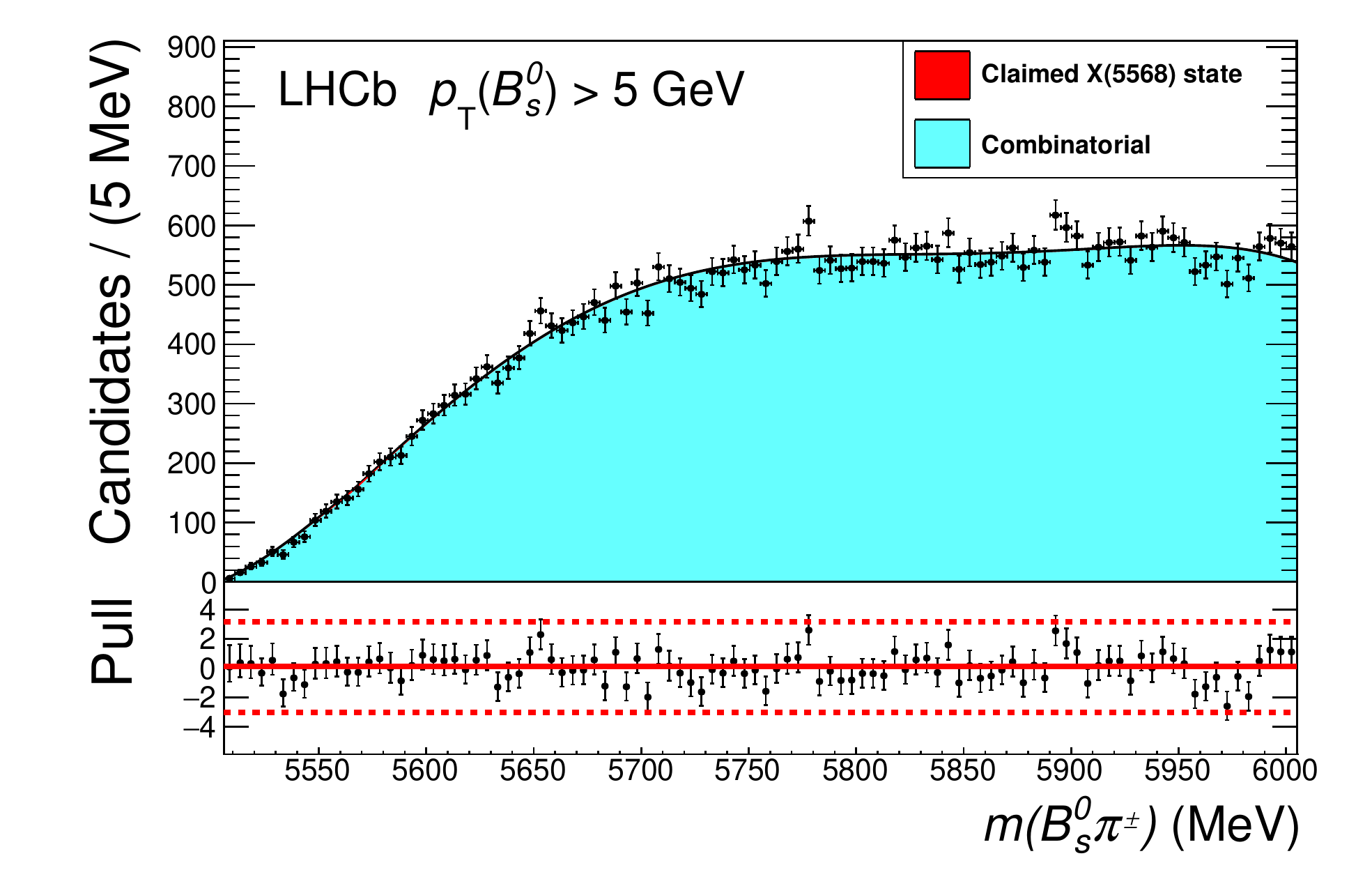}\\
  \includegraphics[width=0.65\textwidth]{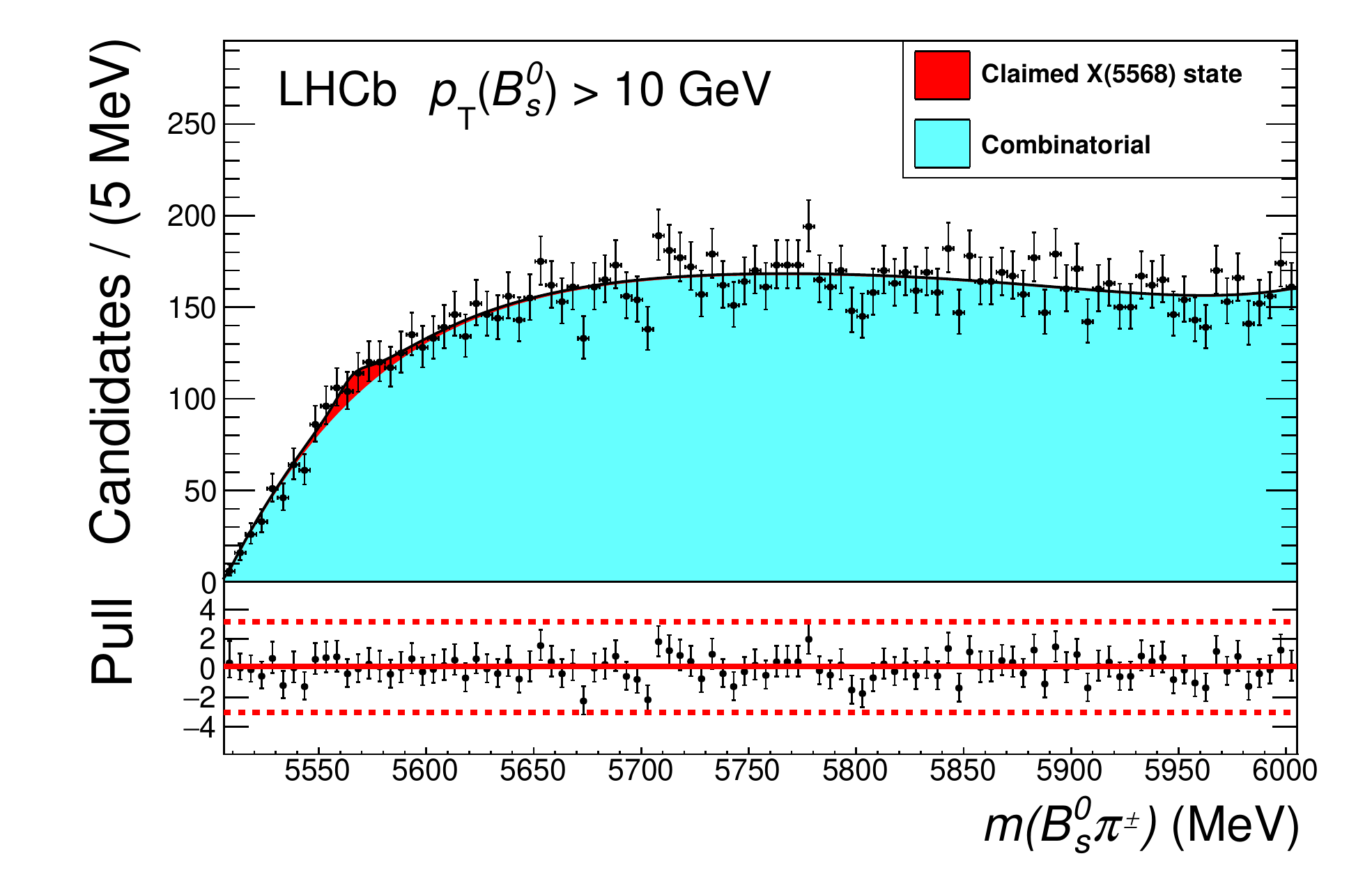}\\
  \includegraphics[width=0.65\textwidth]{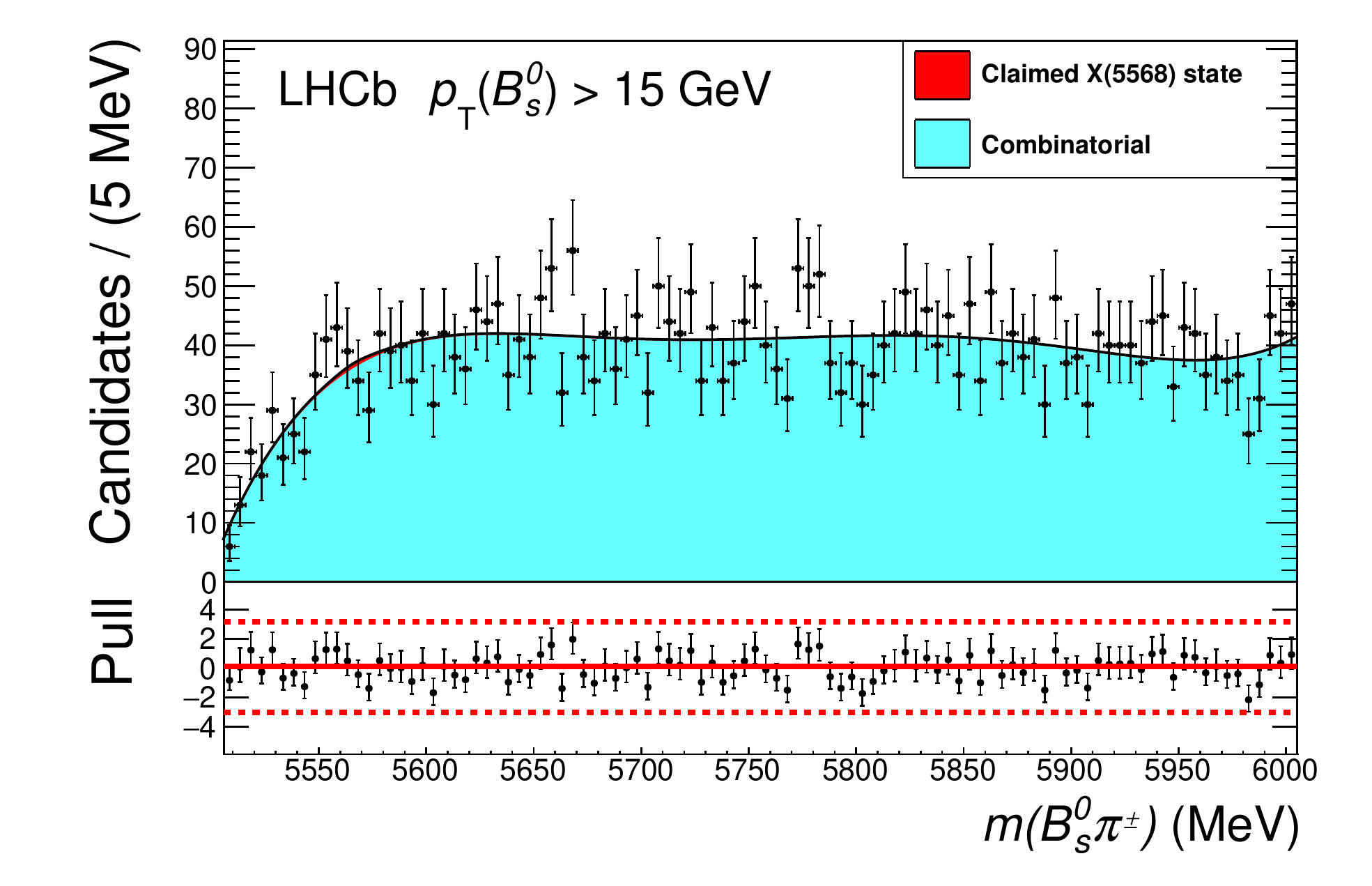}\\
}
  \caption{\small
    Results of the fit to the $\Bs\pipm$ mass distribution for candidates (both \Bs modes combined) with minimum $\pt(\Bs)$ of (top) $5 \gev$, (middle) $10 \gev$ and (bottom) $15 \gev$.
    The component for the claimed $X(5568)$ state is included in the fit but is not significant.
    The distributions of the normalised residuals, or ``pulls'', displayed underneath the main figures show good agreement between the fit functions and the data.
  }
  \label{fig:BsPiFit}
\end{figure}

For each choice of signal mass and width parameters, a binned maximum likelihood fit to the $\Bs\pipm$ candidate mass spectrum is used to determine the signal and background yields and the parameters of the polynomial shape that describes the background.
The two \Bs decay modes are fitted simultaneously.
The results of the fit where the mass and width are fixed according to the central values obtained by the \dzero collaboration, $m = 5567.8 \pm 2.9 \stat \, ^{+0.9}_{-1.9} \syst \mev$ and $\Gamma = 21.9 \pm 6.4 \stat \, ^{+5.0}_{-2.5} \syst \mev$~\cite{D0:2016mwd}, are shown in Fig.~\ref{fig:BsPiFit} for both \Bs decay modes combined.
The $X(5568)$ yield is not significant for any minimum $\pt(\Bs)$ requirement.
In each case the change in negative log-likelihood between fits including or not including the signal component is less than 2 units for two additional free parameters corresponding to the yields in the two \Bs decay modes. 
The results of the fits are summarised in Table~\ref{tab:fitResults}.

The yields $N$ obtained from the fits are used to measure the ratio of cross-sections
\begin{eqnarray}
  \rho_X^{\rm LHCb} & \equiv & \frac{\sigma(pp \to X + {\rm anything})\times {\cal B}\left(X \to \Bs\pipm\right)}{\sigma(pp \to \Bs + {\rm anything})} \, , \\
  & = & \frac{N(X)}{N(\Bs)} \times \frac{1}{\epsilon^{\rm rel}(X)} \, , 
\end{eqnarray}
where the cross-sections $\sigma$ are for promptly produced particles within the LHCb acceptance.
Since $\sigma(pp \to \Bs + {\rm anything})$ in the LHCb acceptance has been previously measured~\cite{LHCb-PAPER-2013-004}, any result for $\rho_X^{\rm LHCb}$ can be scaled to give a result for $\sigma(pp \to X + {\rm anything})\times {\cal B}\left(X \to \Bs\pipm\right)$ in the LHCb acceptance.
The relative efficiency $\epsilon^{\rm rel}(X) = \frac{\epsilon(X)}{\epsilon(\Bs)}$ accounts for the reconstruction and selection efficiency of the companion pion as well as the requirement that it is within the LHCb acceptance.
These effects are determined from simulation, weighted to reproduce the measured differential \Bs production spectrum~\cite{LHCb-PAPER-2013-004}, together with a data-driven evaluation~\cite{LHCb-DP-2012-003} of the efficiency of the particle identification requirement on the companion pion.
In the simulation the $X$ state is assumed to be spinless; it has been verified that the systematic uncertainty associated with this choice is negligible.
The quantities used to evaluate $\rho_X^{\rm LHCb}$ are summarised in Table~\ref{tab:fitResults}.

\begin{table}[tb]
  \centering
  \caption{\small
    Yields, $N$, of \Bs and $X(5568)$ candidates obtained from the fits to the \Bs and $\Bs\pipm$ candidate mass distributions, with statistical uncertainties only.
    The values reported for $N(\Bs)$ are those inside the \Bs signal window.
    The reported values for $X(5568)$ are obtained from fits with signal mass and width parameters fixed to those determined by the D0 collaboration.
    Relative efficiencies $\epsilon^{\rm rel}(X)$ of the \Bs and $X(5568)$ candidate selection criteria are also given. The reported uncertainties on the
    relative efficiencies are only statistical, due to the finite size of the simulated samples.
  }
  \label{tab:fitResults} 
\ifthenelse{\boolean{wordcount}}{}{
  \begin{tabular}{cl r@{$\,\pm\,$}l r@{$\,\pm\,$}l r@{$\,\pm\,$}l}
    \hline \\ [-2.5ex]
    & & \multicolumn{2}{c}{$\Bs\to\Dsm\pip$} & \multicolumn{2}{c}{$\Bs \to \jpsi\phi$} & \multicolumn{2}{c}{\phantom{$-$}\,Sum} \\
    \hline
     & $\pt(\Bs) > \phantom{1}5 \gev$ & 62.2 & 0.3 & 43.6 & 0.2 & 105.8 & 0.4 \\
    $N(\Bs)/10^3$ & $\pt(\Bs) > 10 \gev$ & 28.4 & 0.2 & 13.2 & 0.1 & 41.6 & 0.2 \\
     & $\pt(\Bs) > 15 \gev$ & 8.8 & 0.1 & 3.7 & 0.1 & 12.5 & 0.1 \\
    \hline
     & $\pt(\Bs) > \phantom{1}5 \gev$ & 3 &  64 &  $-$33 & 43 & $-$30 & 77 \\
    $N(X)$ & $\pt(\Bs) > 10 \gev$ & 75 &  52 &  12 & 33 & 87 & 62 \\
     & $\pt(\Bs) > 15 \gev$ & 14 &  31 & $-$10 & 17 & 4 & 35 \\
    \hline
    & $\pt(\Bs) > \phantom{1}5 \gev$ & 0.127 & 0.002 & 0.093 & 0.001 & \multicolumn{2}{c}{\phantom{$-$}\,---} \\ 
    $\epsilon^{\rm rel}(X)$ & $\pt(\Bs) > 10 \gev$ & 0.213 & 0.003 & 0.206 & 0.002 & \multicolumn{2}{c}{\phantom{$-$}\,---} \\
    & $\pt(\Bs) > 15 \gev$ & 0.289 & 0.005 & 0.290 & 0.004 & \multicolumn{2}{c}{\phantom{$-$}\,---} \\
    \hline
  \end{tabular}
}
\end{table}

Systematic uncertainties are assigned due to possible biases in the evaluation of $N(X)$, $N(\Bs)$ and $\epsilon^{\rm rel}(X)$.
The signal shape is modified by varying the efficiency function, and separately by changing the assumed angular momentum in the relativistic Breit--Wigner function from S-wave to P-wave. In each case, the changes in $N(X)$ are assigned as the associated systematic uncertainties.
Uncertainties associated with the determination of $N(\Bs)$ arise due to the size of the \Bs sample and the estimation of the background in the signal region.
In addition to the limited size of the simulation sample, uncertainties associated with $\epsilon^{\rm rel}(X)$ arise due to the precision with which the companion pion reconstruction and particle identification efficiencies are known~\cite{LHCB-DP-2013-002,LHCb-DP-2012-003}.
The uncertainties from different sources are combined in quadrature and give a total that is much smaller than the statistical uncertainty.
To obtain results that can be compared to those for the claimed $X(5568)$ state reported by the D0 collaboration, additional systematic uncertainties are assigned from the changes in the results for $\rho_X^{\rm LHCb}$ when the mass and width parameters are varied independently within $\pm1\sigma$ ranges from their central values.
These are the dominant sources of systematic uncertainty.

To cross-check the results, candidates are selected with criteria similar to those used in the observation of $\Bc \to \Bs\pip$ decays~\cite{LHCb-PAPER-2013-044}, with consistent results.
In addition, $\Bz\to\Dm\pip$ decays are used to create $\Bz\pip$ combinations, and the results on the excited \B states of Ref.~\cite{LHCb-PAPER-2014-067} are reproduced.

The values of $\rho_X^{\rm LHCb}$ for the two \Bs decay modes are consistent and are therefore combined in a weighted average.
In the average, systematic uncertainties are taken to be uncorrelated between the two \Bs decay modes.
An exception is made when obtaining results corresponding to the claimed $X(5568)$ state, where the uncertainty due to the limited precision of the reported mass and width values~\cite{D0:2016mwd} is treated as correlated between the two modes.
These results are
\begin{eqnarray*}
  \rho_X^{\rm LHCb}(\pt(\Bs) > \phantom{1}5 \gev) & = & -0.003 \pm 0.006 \pm 0.002 \, ,\\
  \rho_X^{\rm LHCb}(\pt(\Bs) > 10 \gev) & = & \phantom{-}0.010 \pm 0.007 \pm 0.005 \, ,\\
  \rho_X^{\rm LHCb}(\pt(\Bs) > 15 \gev) & = & \phantom{-}0.000 \pm 0.010 \pm 0.006 \, ,
\end{eqnarray*}
where the first uncertainty is statistical and the second is systematic. 
Since the signal is not significant, upper limits on $\rho_X^{\rm LHCb}$ are obtained by integration of the likelihood in the positive region to find the value that contains the fraction of the integral corresponding to the required confidence level (CL).
The upper limits at 90\,(95)\,\% CL are found to be 
\begin{eqnarray*}
  \rho_X^{\rm LHCb}(\pt(\Bs) > \phantom{1}5 \gev) & < & 0.011 \ (0.012) \, ,\\
  \rho_X^{\rm LHCb}(\pt(\Bs) > 10 \gev) & < & 0.021 \ (0.024) \, ,\\
  \rho_X^{\rm LHCb}(\pt(\Bs) > 15 \gev) & < & 0.018 \ (0.020) \, . 
\end{eqnarray*}

\begin{figure}[!tb]
\ifthenelse{\boolean{wordcount}}{}{
  \centering
  \includegraphics[width=0.85\textwidth]{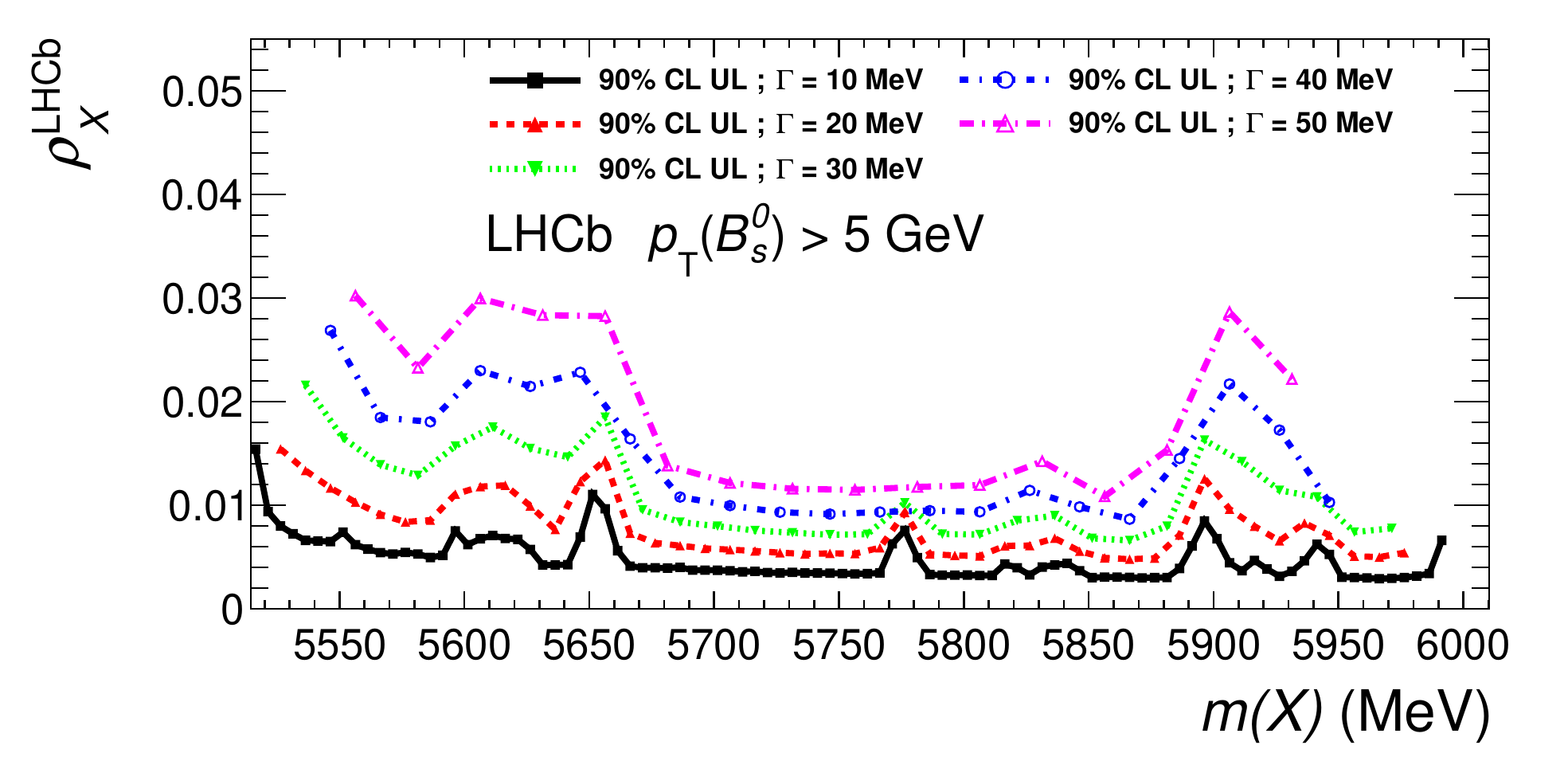}\\
  \includegraphics[width=0.85\textwidth]{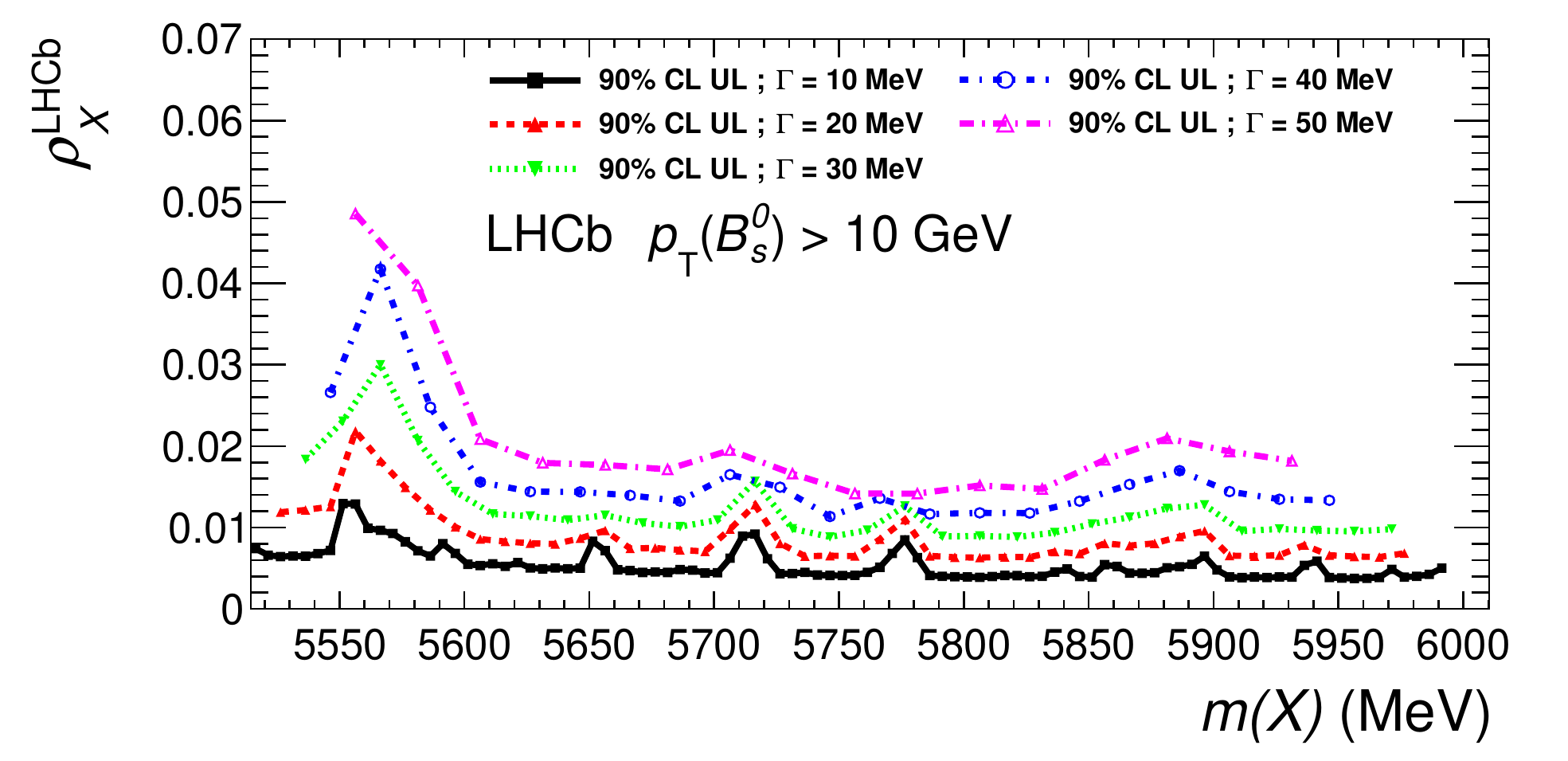}\\
  \includegraphics[width=0.85\textwidth]{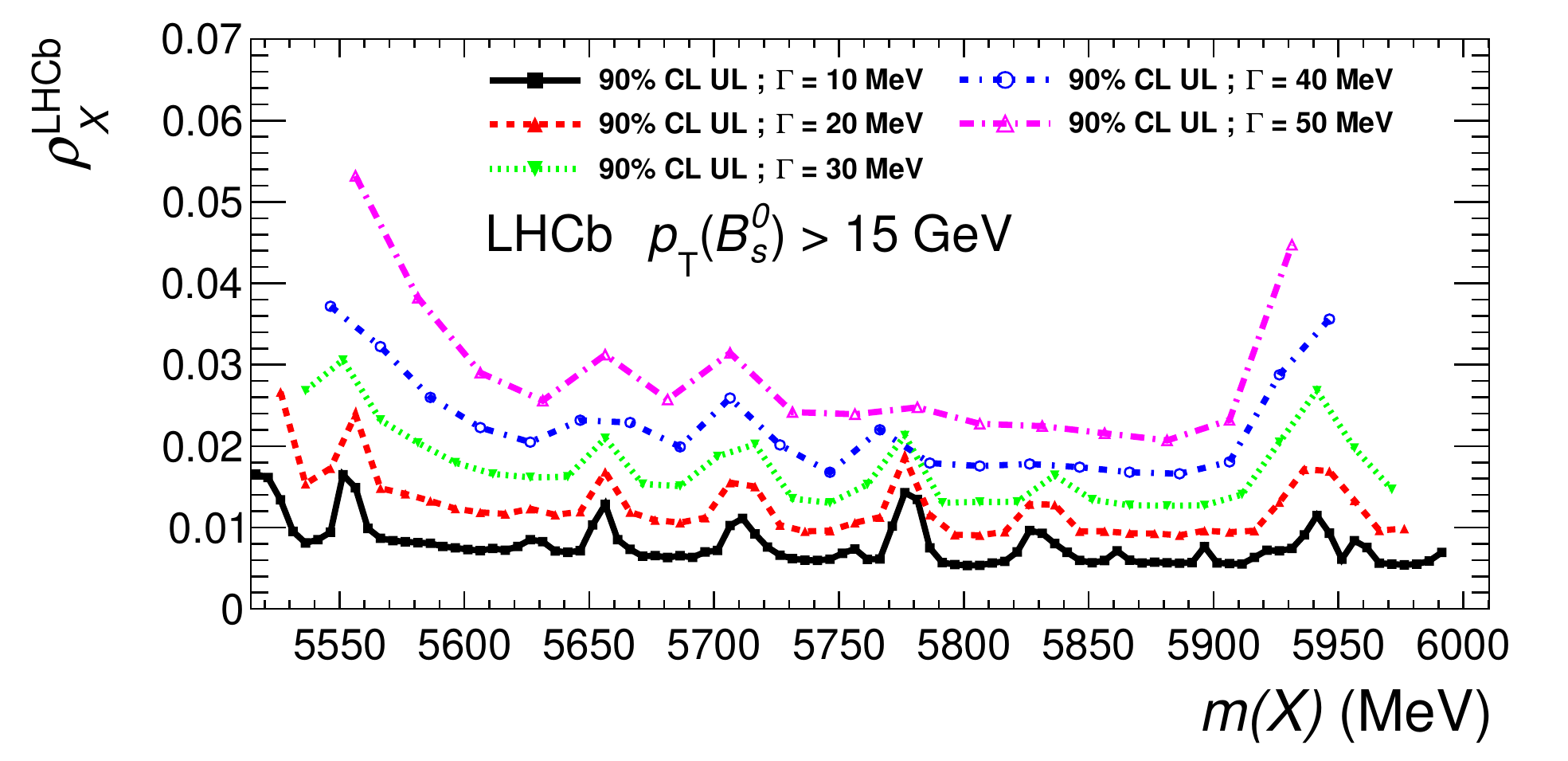}\\
}
  \caption{\small 
    Upper limits (ULs) at 90\,\% confidence level (CL) as functions of the mass and width of a purported exotic state $X$ decaying to $\Bs\pipm$ with minimum $\pt(\Bs)$ of (top) $5 \gev$, (middle) $10 \gev$ and (bottom) $15 \gev$. The same limits also apply to a possible exotic meson decaying through the chain $\Bss\pipm$, $\Bss \to \Bs \gamma$ where the photon is excluded from the reconstructed decays. In the latter case the nominal mass difference $m(\Bss) - m(\Bs) = 48.6^{+1.8}_{-1.6} \mev$~\cite{PDG2014} has to be added to the values on the x-axis to get the mass of the exotic meson under investigation.
  }
  \label{fig:ULs}
\end{figure}

No significant signal for a $\Bs\pipm$ resonance is seen at any value of mass and width in the range considered.
To obtain limits on $\rho_X^{\rm LHCb}$ for different values of these parameters, fits are performed for widths ($\Gamma$) of $10$ to $50 \mev$ in $10 \mev$ steps.
For each width, the mass is scanned in steps of $\Gamma/2$, starting one unit of width above the kinematic threshold and ending approximately one and a half units of width below $6000 \mev$.
The upper edge of the range is chosen because an exotic state with higher mass would be expected to give a clearer signature in the $\Bz\Kpm$ final state~\cite{Esposito:2016itg}.
The results are obtained in the same way as described above, and converted into upper limits that are shown in Fig.~\ref{fig:ULs}.
The upper limits are weaker when a broader width is assumed, due to the larger amount of background under the putative peak.
The limits also become weaker when there is an excess of events in the signal region, although all such excesses are consistent with being statistical fluctuations. 
The method used to set the upper limits smooths out any negative fluctuations.

In summary, a search for the claimed $X(5568)$ state has been carried out using a data sample corresponding to $3~\invfb$ of $pp$ collision data at $\sqrt{s} = 7$ and $8~\tev$ recorded by LHCb.
No significant excess is found and thus the existence of the $X(5568)$ state is not confirmed.
Upper limits are set on the relative production rate of the claimed state in the LHCb acceptance.
Limits are also set as a function of the mass and width of a possible exotic meson decaying to the $\Bs\pipm$ final state.
The same limits also apply to a possible exotic meson decaying through the chain $\Bss\pipm$, $\Bss \to \Bs \gamma$ where the photon is excluded from the reconstructed decays.



\section*{Acknowledgements}

\noindent We express our gratitude to our colleagues in the CERN
accelerator departments for the excellent performance of the LHC. We
thank the technical and administrative staff at the LHCb
institutes. We acknowledge support from CERN and from the national
agencies: CAPES, CNPq, FAPERJ and FINEP (Brazil); NSFC (China);
CNRS/IN2P3 (France); BMBF, DFG and MPG (Germany); INFN (Italy); 
FOM and NWO (The Netherlands); MNiSW and NCN (Poland); MEN/IFA (Romania); 
MinES and FASO (Russia); MinECo (Spain); SNSF and SER (Switzerland); 
NASU (Ukraine); STFC (United Kingdom); NSF (USA).
We acknowledge the computing resources that are provided by CERN, IN2P3 (France), KIT and DESY (Germany), INFN (Italy), SURF (The Netherlands), PIC (Spain), GridPP (United Kingdom), RRCKI and Yandex LLC (Russia), CSCS (Switzerland), IFIN-HH (Romania), CBPF (Brazil), PL-GRID (Poland) and OSC (USA). We are indebted to the communities behind the multiple open 
source software packages on which we depend.
Individual groups or members have received support from AvH Foundation (Germany),
EPLANET, Marie Sk\l{}odowska-Curie Actions and ERC (European Union), 
Conseil G\'{e}n\'{e}ral de Haute-Savoie, Labex ENIGMASS and OCEVU, 
R\'{e}gion Auvergne (France), RFBR and Yandex LLC (Russia), GVA, XuntaGal and GENCAT (Spain), Herchel Smith Fund, The Royal Society, Royal Commission for the Exhibition of 1851 and the Leverhulme Trust (United Kingdom).

  \addcontentsline{toc}{section}{References}
  \bibliography{main,references,LHCb-PAPER,LHCb-CONF,LHCb-DP,LHCb-TDR}
 
\ifthenelse{\boolean{wordcount}}{}{ 
  \clearpage
  \centerline{\large\bf LHCb collaboration}
\begin{flushleft}
\small
R.~Aaij$^{40}$,
B.~Adeva$^{39}$,
M.~Adinolfi$^{48}$,
Z.~Ajaltouni$^{5}$,
S.~Akar$^{6}$,
J.~Albrecht$^{10}$,
F.~Alessio$^{40}$,
M.~Alexander$^{53}$,
S.~Ali$^{43}$,
G.~Alkhazov$^{31}$
% ...

{\footnotesize \it
% ...
$ ^{5}$Clermont Universit{\'e}, Universit{\'e} Blaise Pascal, CNRS/IN2P3, LPC, Clermont-Ferrand, France\\
$ ^{6}$CPPM, Aix-Marseille Universit{\'e}, CNRS/IN2P3, Marseille, France\\
% ...
$ ^{10}$Fakult{\"a}t Physik, Technische Universit{\"a}t Dortmund, Dortmund, Germany\\
% ...
$ ^{31}$Petersburg Nuclear Physics Institute (PNPI), Gatchina, Russia\\
% ...
$ ^{39}$Universidad de Santiago de Compostela, Santiago de Compostela, Spain\\
$ ^{40}$European Organization for Nuclear Research (CERN), Geneva, Switzerland\\
% ...
$ ^{43}$Nikhef National Institute for Subatomic Physics, Amsterdam, The Netherlands\\
% ...
$ ^{48}$H.H. Wills Physics Laboratory, University of Bristol, Bristol, United Kingdom\\
% ...
$ ^{53}$School of Physics and Astronomy, University of Glasgow, Glasgow, United Kingdom\\
% ...
}
\end{flushleft}
}

\end{document}